# Pressure-Induced Confined Metal from the Mott Insulator $Sr_3Ir_2O_7$

Yang Ding,[1,2,3*] Liuxiang Yang,[1,3] Cheng-Chien Chen,[1,4] Heung-Sik Kim,[5] Myung Joon Han,[5] Wei Luo,[6] Zhenxing Feng,[7] Mary Upton,[2] Diego Casa,[2] Jungho Kim,[2] Thomas Gog,[2] Zhidan Zeng,[1,3] Gang Cao,[8] Ho-kwang Mao,[1, 3, 9] and Michel van Veenendaal[2, 10]

[1] Center for High Pressure Science and Technology Advanced Research, Pudong, Shanghai 201203, China

[2] Advanced Photon Source, Argonne National Laboratory, Argonne, Illinois 60439, USA

[3] HPSynC, Geophysical Laboratory, Carnegie Institution of Washington, Argonne, Illinois 60439, USA

[4] Department of Physics, University of Alabama at Birmingham, Birmingham, Alabama 35294, USA

[5] Department of Physics, Korea Advanced Institute of Science and Technology, Daejeon 305-701, Republic of Korea

[6] Condensed Matter Theory Group, Department of Physics Box 530, SE-751 21 Uppsala, Sweden

[7] Chemical Sciences and Engineering, Argonne National Laboratory, Argonne, Illinois 60439, USA

[8] Department of Physics and Astronomy, University of Kentucky, Lexington, Kentucky 40506, USA

[9] Geophysical Laboratory, Carnegie Institution of Washington, Washington, D. C. 20015, USA

[10] Department of Physics, Northern Illinois University, De Kalb, Illinois 60115, USA

*yang.ding@hpstar.ac.cn

Jan 30, 2016

LK15397

**The spin-orbit Mott insulator $Sr_3Ir_2O_7$ provides a fascinating playground to explore insulator-metal-transition driven by intertwined charge, spin, and lattice degrees of freedom. Here we report high-pressure electric resistance and resonant inelastic X-ray scattering measurements on single-crystal $Sr_3Ir_2O_7$ up to 63-65 GPa at 300 K. The material becomes a confined metal at 59.5 GPa, showing metallicity in the *ab*-plane but an insulating behavior along the *c*-axis. Such unusual phenomenon resembles the strange metal phase in cuprate superconductors. Since there is no sign of the collapse of spin-orbit or Coulomb interactions in X-ray measurements, this novel insulator-metal transition is potentially driven by a first-order structural change at nearby pressures. Our discovery points to a new approach for synthesizing functional materials.**

The 5$d$ transition-metal oxides have recently attracted tremendous interest because the interplay between spin-orbit interaction and electron correlation in these materials can lead to novel states of matter, such as quantum spin liquids, topological orders, and potentially high-temperature superconductors [1-5]. Among these compounds, the Ruddlesden-Popper series of iridates Sr$_{n+1}$Ir$_n$O$_{3n+1}$ (where $n$ is the number of SrIrO$_3$ perovskite layers between extra SrO layers) has been extensively studied. In these systems, the nearly cubic crystal field splits the 5$d$ shell into the $e_g$ and $t_{2g}$ levels leading to a $t^5_{2g}$ configuration for Ir$^{4+}$. The strong spin-orbit interaction further splits the $t_{2g}$ orbitals into $J_{\text{eff}}$ = 1/2 and 3/2 states. The singly-occupied $J_{\text{eff}}$ = 1/2 state can be further split by the electron-electron Coulomb repulsion, giving rise to Mott insulating behavior [6, 7]. With their correlated spin, charge, and lattice responses to external perturbation, perovskite iridates provide a fascinating system for studying the competition and cooperation of fundamental interactions.

The double-layered perovskite Sr$_3$Ir$_2$O$_7$ [8] appears in a unique position in this series [4, 8-15]. Its structure has been regarded to be tetragonal (*I4/mmm*, with $a$ = 3.9026 Å and $c$ = 20.9300 Å [16, 17]), but other studies also reveal that the crystal symmetry could be orthorhombic (*Bbcb*, with $a$ = 5.522 Å, $b$ = 5.521 Å, and $c$ = 20.917 Å [18]) or even monoclinic (*C2/c*, with $a$ = 5.5185 Å, $b$ = 5.5099 Å, and $c$ = 20.935 Å [19]). The charge gap of Sr$_3$Ir$_2$O$_7$ (~ 0.1-0.3 eV) is comparable to the magnitude of the underlying exchange interaction [20-22], and such an insulating state is considered in the weak Mott limit. Small perturbations, such as carrier doping, magnetic field, or external pressure, can affect the stability of the insulating phase. Therefore, this system provides a valuable platform for exploring the collapse of the Mott gap. Indeed, in addition to a dimensionality-driven insulator-metal transition (IMT) in the iridate Ruddlesden-Popper series [23], recent experiments have found that a small amount of La substitution for Sr in (Sr$_{1-x}$La$_x$)$_3$Ir$_2$O$_7$ can melt away its insulating gap and lead to a correlated metallic state [24], showing that the system resides in close proximity to an IMT phase boundary.

Conventional wisdom would dictate that pressure has a similar effect as carrier doping. However, to date, high-pressure studies on $Sr_3Ir_2O_7$ have been inconclusive due to the lack of consistent experimental results. Li et al. claimed that $Sr_3Ir_2O_7$ undergoes an IMT-like transition at ~13 GPa [25], but Zocco et al. found that the insulating state persists up to 104 GPa [12]. In addition, whereas Zhao et al. reported a second-order structure transition at ~15 GPa, Donnere et al. only discovered a first-order structure transition at ~54 GPa [17]. Based on first-principles calculations, Donnere et al. also predicted an IMT concurring with the structural change [17]. Further experiments are thereby necessary to clarify these ambiguities.

In this Letter, we present electric resistance and resonant inelastic x-ray scattering (RIXS) [26] measurements for the first time on single-crystal $Sr_3Ir_2O_7$ in a diamond anvil cell (DAC) at pressures up to ~63-65 GPa at 300 K. The resistance measurements indicate that an IMT is present at ~59.5 GPa, with the high-pressure phase exhibiting a novel confinement phenomenon: a metallic behavior within the *ab*-plane but an insulating one along the *c*-axis. This intriguing phenomenon is reminiscent of the strange metal phase in cuprate superconductors [27, 28]. Because the RIXS measurement shows robust spin-orbit and electron Coulomb interactions within the investigated pressure range, this novel IMT is potentially driven by a first-order structure change at a nearby pressure as reported by recent diffraction measurements.[17]

Figure 1 shows the electric resistance measurements on a $Sr_3Ir_2O_7$ single crystal at different pressures. In this experiment, resistances within the *ab*-plane $R_{ab}$ (Fig. 1a) and along the *c*-axis $R_c$ (Fig. 1b) are measured at the same time for each pressure point with the standard four-probe-electrode-circuit method [29]. Further experimental details are given in the *Supplementary*

*Information* [30]. At low pressures (18.4 and 43.15 GPa), the system is an insulator, with its resistances increasing with decreasing temperature ($dR_{ab}/dT < 0$ and $dR_c/dT < 0$). However, at 59.5 and 63.0 GPa (Fig. 1d), the material shows a clear metallic behavior within the *ab*-plane ($dR_{ab}/dT > 0$) and an insulating-like state along the *c*-axis ($dR_c/dT < 0$). Such behavior is highly unusual. Although the resistivity ratio between the *ab*-plane and *c*-axis could differ by several orders of magnitude for an anisotropic metal, localization in only one direction rarely occurs. This novel "confinement" phenomenon is reminiscent of the strange metal phase in over-doped cuprate superconductors [27, 28]. Together with the recent discovery of Fermi-arc features in $Sr_2IrO_4$, the similarity between the high-pressure phase of $Sr_3Ir_2O_7$ and the cuprates shows the potential promise of superconductivity in perovskite iridates $Sr_{n+1}Ir_nO_{3n+1}$ [35, 36]. To the best of our knowledge, this is the first time that a confined metal is found at high pressure. Our discovery thereby points to a new way of synthesizing novel states of matter.

IMTs in correlated transition-metal oxides could be driven by bandwidth broadening, structural transition, or the collapse of electron-electron interaction.[37, 38] Furthermore, the insulating state in $Sr_3Ir_2O_7$ is associated with a strong spin-orbit interaction, another parameter that can trigger an IMT. An understanding of the mechanism underlying the pressured-induced IMT in $Sr_3Ir_2O_7$ requires a detailed knowledge of the pressure evolution of the electronic structure. However, such information was unattainable until our recent integration of the RIXS and DAC techniques. [39]

Figure 2 displays the high-pressure Ir $L_3$-edge RIXS data collected with a scattering vector along the [0, $\pi/b$, 0] (indexed by a pesudo-tetragonal symmetry for simplicity) direction in a horizontal scattering geometry. In the high-pressure experiments, $Sr_3Ir_2O_7$ single crystal is loaded with neon gas as the pressure medium in a Mao-type symmetric DAC. Based on diffraction measurements [17], the unit cell of $Sr_3Ir_2O_7$ has been compressed by nearly 6% at 64.6 GPa.

Accordingly, from 0.98 to 64.8 GPa, the scattering vector $q$ spans from $[0,\pi,0]$ at 0.98 GPa, $[0,0.99\pi,0]$ at 12.4 GPa, $[0,0.96\pi,0]$ at 23.6 GPa, $[0,0.95\pi,0]$ at 34.4 GPa, and to $[0,0.94\pi,0]$ at 43.8 GPa, due to a sample size reduction with pressure [see the *Supplementary Information for further details* [30]. In addition to the zero-energy quasielastic peak (indicated as A), three other major peaks in the RIXS map (Fig. 2a) were identified at energy loss of 0.25 (B), 0.64 (C), and 2.5 (D) eV. The resonant energy for peak D is 11.218 KeV, and the resonant energy for C and B is 11.216 KeV.

Previous RIXS studies [40-43] have identified peak D as a crystal-field excitation between the Ir $5d$ $t_{2g}$ and $e_g$ orbitals, possibly mixed with a small amount (5%) of charge-transfer excitation from the oxygen $2p$ ligands [44]. In a nearly cubic crystal field, peak C is the spin-orbit exciton related to transition between the $J_{\text{eff}} =1/2$ and 3/2 states [35, 41, 45]. In other iridate compounds, a non-cubic splitting of $J_{\text{eff}}=1/2$ was also observed [35, 42, 43, 46]. In the low-energy region (< 0.3 eV), peak B has been previously attributed to excitations of magnetic origin [47-50], which may overlap with phonon excitations of the DAC and the sample. No significant change is observed in the low-energy area of peaks A and B during our measurements.

To more accurately obtain the energy loss and line width of peaks C and D, a peak-fitting analysis is performed. The results are shown in Figs. 3a and 3b, as well as in Table I. Figure 3 displays two significant changes in peak D: First, whereas its excitation energy grows with applied pressure, it is only weakly pressure-dependent above 53.5 GPa (Fig. 3a). Second, the line width drops significantly above 53.5 GPa (Fig. 3b). In comparison, the energy of peak C increases with pressure up to 23.6 GPa, then decreases. Above 53.5 GPa, peak C's energy is increased by ~15% compared to that at ambient condition (Fig. 3a). Overall, the line width of peak C is increased by pressure, but significantly reduced above 53.5 GPa. Although a tetragonal splitting could increase the line width of peak C, this change also should have been found in peak D (this change would be anticipated in the peak D as well but is not observed, as shown in Fig.3b); yet Fig. 3b does not show such a change. Therefore, the different pressure dependence between peak D and peak C below 53.5 GPa could imply that a weak dispersion of spin-orbiton excitation exists in the low pressure phase, despite that the excitation has a predominant intra-site character [45, 46]. In RIXS measurements of $Sr_2IrO_4$, the dispersive behavior of spin-orbiton excitation is also more obvious than the crystal field excitation [40]. Above 53.5 GPa, the pressure dependence of peak C and peak D becomes more complicated.

As the line widths of peaks C and D both drop above 53.5 GPa when the $ab$-plane of $Sr_3Ir_2O_7$ becomes metallic, a correlation appears between the line width reduction and the IMT. Because the $IrO_6$ octahedron symmetry becomes even lower above 54 GPa [17], the line width reduction cannot arise from the tetragonal splitting. In contrast, a self-energy correction due to electron interaction could contribute to the intrinsic lifetime broadening. Therefore, the sudden drop of the line width for both peaks C and D could indicate that the effective Coulomb repulsion $U$ is slightly weakened in the high-pressure metallic phase due to enhanced screening. Nonetheless, because the bandwidth broadening and lattice changes are expected to be more substantial, the small reduction of effective $U$ should be regarded as a consequence, instead of a cause, of the pressure-induced IMT.

In addition, the RIXS results unambiguously show that spin-orbit coupling strength $\lambda$ (equal to two thirds the energy of peak C) is robust up to 64.8 GPa, where the system already becomes metallic according to our resistance measurements. Therefore, the pressure-induced IMT in $Sr_3Ir_2O_7$ cannot be driven by a collapse of the spin-orbit interaction. We also note that although the $L_2/L_3$-edge branching ratio in X-ray absorption spectroscopy could be used to obtain the ground state expectation value of $<L\cdot S>$, the $\lambda$ value alone cannot be determined in this manner [51, 52].

We now discuss further implications of our RIXS data by comparing them to diffraction measurements. If the energy loss of peak D is approximated as the crystal-field splitting $10Dq$ in a perfect $O_h$ symmetry, we can examine whether the relationship $10Dq = K/d^n_{\text{Ir-O}}$ ($n$ = 3.5-7) is still valid under pressure, where $K$ is a constant and $d_{\text{Ir-O}}$ is the Ir-O bond length. However, Fig. 3c shows that $10Dq$ (represented by peak D's energy loss) and the effective Ir-O bond length (represented by the lattice parameter $a$) no longer comply with the relation $10Dq = K/d^n_{\text{Ir-O}}$ ($n$= 3.5-7). This result suggests that the *ab*-plane Ir-O-Ir bond angle $\phi$ must be taken into account.[53, 54] In this case, the symmetry of $Sr_3Ir_2O_7$ in the low-pressure, insulating state cannot be effectively treated as *I4/mmm* [16, 17]. This conclusion is supported by earlier measurements that determined the crystal symmetry of $Sr_3Ir_2O_7$ as orthorhombic (space group *Bbca*) [18, 25] or even monoclinic (space group *C2/c*) [12] with $\phi$ equal to 158° at ambient conditions.

By using $\phi$ and $n$ as variation parameters, we obtained the Ir-O bond length and bond angle using $10Dq = K/d^n_{\text{Ir-O}}$ (with details given in the *Supplementary Information* [30]). The results are plotted in Fig. 3d and listed in Table II. The best fit is obtained for $n$=4.6, which is comparable to the results in other 5$d$ transition-metal oxides [55]. Figure 3d shows that the bond angle $\phi$ decreases when the system is approaching the metallic phase below the critical pressure, thereby leading to an increased rotation of $IrO_6$ octahedron with respect to the *c*-axis. In fact, such a conclusion remains qualitatively valid for all $n$ = 3.5-7 in the fitting. Such a pressure-induced octahedron rotation could provide a de-hybridization mechanism that prevents the system from becoming metallic [56, 57] at pressures below 50 GPa.

As a first-order approximation, the pervoskite structure tolerance factor,[58] $t = \frac{R_{Sr} + R_o}{\sqrt{2}(R_{Ir} + R_o)}$ (where $R_{Sr}$, $R_o$, and $R_{Ir}$ represent the atomic radii of Sr, O, and Ir, respectively), can be used to gauge the stability of structure. When the rotation angle is considered, the tolerance factor $t$ is proportional to $1/\sin(\phi/2)$. Based on our fitting results, $t$ reaches 1.02-1.04 when the rotation angle becomes less than 152° at the critical pressure of 54 GPa as reported by diffraction measurements.

This value reaches the $t =1.04$ perovskite stability limit [59]. Therefore, the structural transition above 54 GPa is likely triggered by a saturated $IrO_6$ octahedron rotation. The first-order structural change in turn could drive the pressure-induced insulator-metal transition in $Sr_3Ir_2O_7$, since there are no apparent changes in the electron Coulomb repulsion and spin-orbit interactions up to 65 GPa as revealed by our RIXS measurements.

Finally, since the critical pressure (53.5 GPa) in RIXS measurements is nearly identical to the structural transition pressure (54 GPa) in diffraction measurements, [17] we believe that the electronic and structural transitions are likely to occur at a nearby pressure, or even concur at the same pressure point. The difference between the structural transition pressure and the insulator-metal-transition pressure (59.5 GPa) determined from our transport measurements is 5.5 GPa. This mismatch is potentially caused by nonhyrostaticity of the Si oil and the coarse pressure step size. We also note that previous transport measurements on powder samples may suffer from uncertainties possibly due to the existence of grain boundaries. Moreover, our single-crystal measurements do not show any abrupt change in the pressure range below 50 GPa, thereby excluding the existence of a structural phase transition below this pressure point, which is consistent with the more recent diffraction measurement by Donnere *et al.* [17].

To conclude, our findings have unraveled a non-trivial interplay between the structural and electronic properties in $Sr_3Ir_2O_7$, and even more generally in the Ruddlesden-Popper series of iridates, where various material families also exhibit correlated spin, charge, and lattice degrees of freedom. Our discovery of a novel high-pressure metallic phase with intriguing confinement phenomenon similar to that found in over-doped cuprate superconductors suggests that superconductivity could potentially be found in doped $Sr_3Ir_2O_7$ under high pressure. Moreover, pressure shows a distinct effect compared to doping or dimensionality-change, thereby pointing to a new way of synthesizing novel states of matter that are inaccessible with other methods.

**ACKNOWLEDGMENTS**

The authors acknowledge useful discussion with Elizabeth Nowadnick. The RIXS measurements are performed at sectors 30 ID-B and 27 ID-B of Advanced Photon Source, a U.S. Department of Energy (DOE) Office of Science User Facility operated by Argonne National Laboratory (ANL) supported by the U.S. DOE Contract No. DE-AC02-06CH11357. C.C.C. is supported by the Aneesur Rahman Postdoctoral Fellowship at ANL. H.-S.K. and M.J.H. were supported by were supported by Basic Science Research Program through NRF (2014R1A1A2057202) and by Samsung Advanced Institute of Technology (SAIT). G.C. acknowledges NSF support via Grant DMR1265162. H.-k. M. acknowledges the support of DOE-BES under Award No. DE-FG02-99ER45775 and NSFC Grant No. U1530402. M.v.V. is supported by the U.S. DOE under Award No. DE-FG02-03ER46097, and by the Institute for Nanoscience, Engineering and Technology at Northern Illinois University.

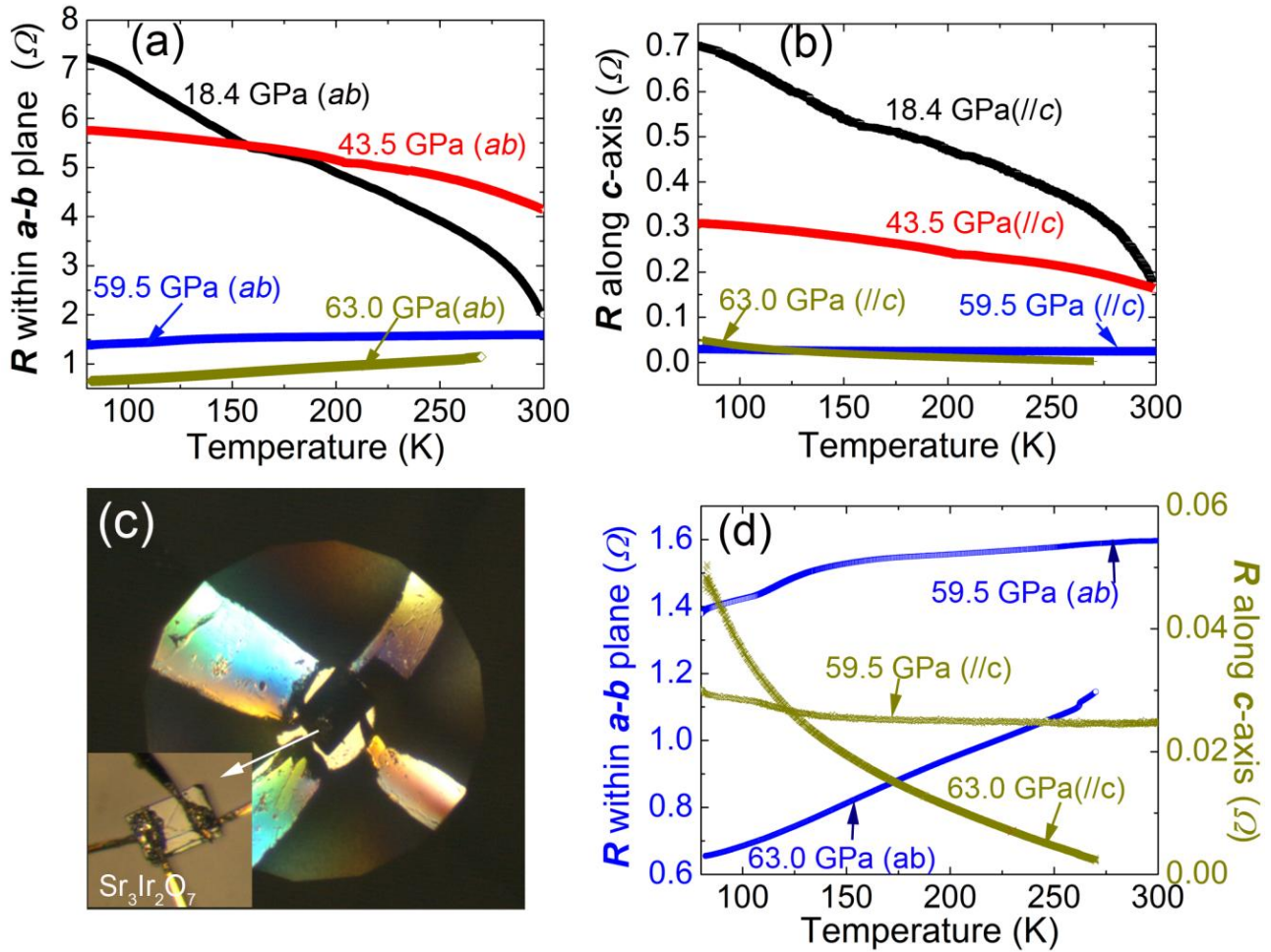

Figure 1. Electric resistance measurements of single crystal $Sr_3Ir_2O_7$ at high pressure (a) within *ab*-plane and (b) along crystal *c*-axis. (c) Four gold electric leads and single crsytal $Sr_3Ir_2O_7$ loaded into a symmetric diamond anvil cell. Inset shows two of the leads attached to the top of the single crystal, and the other two attached to the bottom. Each diamond culet is approximately 300 $\mu$m wide. (d) Temperature dependences of the resistances at 59.5 and 63.0 GPa. The system shows metallic behavior in the *ab*-plane, whereas it still exhibits an insulating state along the *c*-axis. [Authors: this is a single-column figure.]

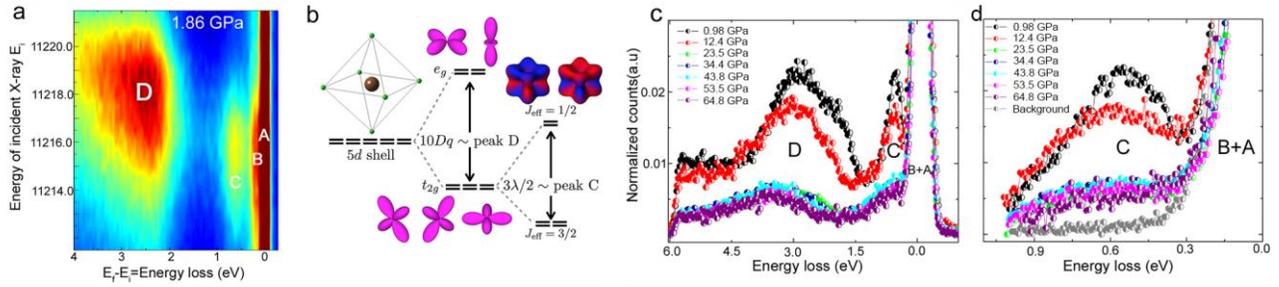

Figure 2. (a) Ir $L_3$-edge RIXS map collected at 1.86 GPa. (b) Atomic energy-level diagram for an $Ir^{4+}$ ion in an octahedral crystal field. (c)-(d) RIXS data collected at high pressures between energy loss of -1.0 to 6.0 eV, and between 0.0 to 1.0 eV, respectively. The incident energy is 11.216 KeV. [Authors: this is a double-column figure.]

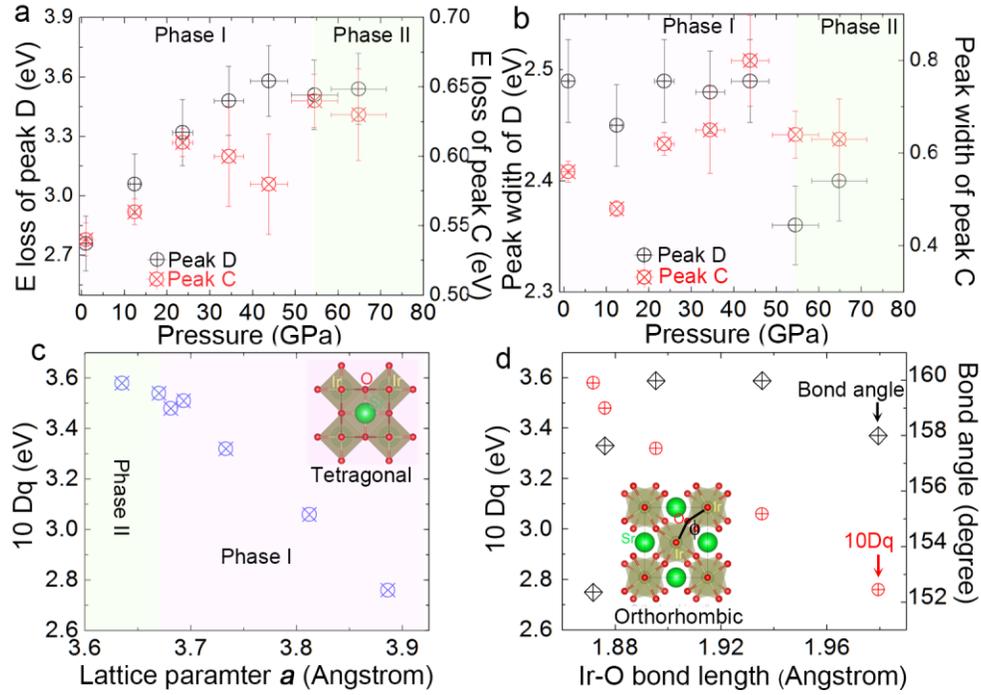

Figure 3. (a) Energies of crystal-field excitation 10$Dq$ measured from peak D and the spin-orbit exciton peak C in RIXS as a function of pressure. (b) Pressure dependences of the RIXS spectral line widths of peaks D and C. (c) Crystal-field excitation energy 10$Dq$ versus the lattice parameter $a$ from diffraction measurements. Areas of different color represent phases with different structures. (d) Fitted $ab$-plane Ir-O bond length the Ir-O-Ir bond angle. Inset displays the orthorhombic structure of $Sr_3Ir_2O_7$ viewed along the $c$-axis in the low-pressure, insulating phase. [Authors: this is a single-column figure.]

Table I. The position and width (full width at half maximum, FWHM) of the crystal-field (CF) excitation peak C and the spin-orbit (SO) exciton peak D obtained from peak fitting analysis. The lattice parameter in the high-pressure monoclinic phase is averaged over the *a*- and *b*-axes.

| P(GPa) | a(Å) | 10Dq(eV) | Width(CF) | SO-exciton (eV) | Width(SO) |
|---|---|---|---|---|---|
| 0.98 | 3.8860 | 2.76 | 2.49 | 0.54 | 0.56 |
| 12.40 | 3.8120 | 3.06 | 2.45 | 0.56 | 0.48 |
| 23.60 | 3.7330 | 3.32 | 2.49 | 0.61 | 0.62 |
| 34.40 | 3.6810 | 3.48 | 2.48 | 0.60 | 0.65 |
| 43.80 | 3.6350 | 3.58 | 2.49 | 0.58 | 0.80 |
| 53.50 | 3.6931 | 3.51 | 2.36 | 0.64 | 0.64 |
| 64.80 | 3.6699 | 3.54 | 2.40 | 0.63 | 0.63 |

Table II. The fitted *ab*-plane Ir-O bond length $d_{\text{Ir-O}}$ and bond angle $\phi$ based on $10Dq = K/d^n_{\text{Ir-O}}$.

|  | P(GPa) | a(Å) | 10Dq(eV) | $\phi$ | $d_{\text{Ir-O}}$ |
|---|---|---|---|---|---|
|  | 0.98 | 3.886 | 2.76 | 158 | 1.9794 |
|  | 12.4 | 3.812 | 3.06 | 159.98 | 1.9355 |
| **n=4.6** | 23.6 | 3.733 | 3.32 | 159.98 | 1.8954 |
|  | 34.4 | 3.681 | 3.48 | 157.64 | 1.8761 |
|  | 43.8 | 3.635 | 3.58 | 152.36 | 1.8717 |